\documentclass[aps,nofootinbib,longbibliography,prd,twocolumn]{revtex4-1}
\usepackage[babel]{csquotes}
\usepackage{graphicx}
\usepackage{amsmath,amssymb}
\usepackage[colorlinks,citecolor=blue,linkcolor=blue,urlcolor=blue]{hyperref}
\usepackage{mathrsfs}
\usepackage{enumerate}
\def\be{\begin{equation}}
\def\ee{\end{equation}}

\newcommand{\cH}{{\cal H}}
\newcommand{\cA}{{\cal A}}

\begin{document}

\title{The subtle unphysical hypothesis of the firewall theorem}

\author{Carlo Rovelli \vspace{1mm}}

\affiliation{\small\mbox{CPT, Aix-Marseille Universit\'e, Universit\'e de Toulon, CNRS, F-13288 Marseille, France,\\ 
and}
}
\date{\small\today}

\begin{abstract}
\noindent The black-hole firewall theorem derives a suspicious consequence (large energy-momentum density at the horizon of a black hole) from a set of seemingly reasonable hypotheses.   I point out the hypothesis which is likely to be unrealistic---a hypothesis not always sufficiently made explicit--- and discuss the subtle confusion at its origin: mixing-up of two different notions of entropy and misusing the entropy bound. 
\end{abstract}
\maketitle

What happens at the horizon of a real black hole?  What would you experience if you crossed it with a starship? We should be able to give a plausible answer to this question using established physics, since a horizon can be a very low-curvature region and there is no obvious reason for unknown physics (like quantum gravity) to play a major role there.   

Yet, today theoretical physicists disagree on the answer.  On the one hand, general relativity suggests that nothing particularly remarkable should happen in a sufficiently small region around a point on the horizon.  On the other hand, a portion of the theoretical physics community convinced itself that the starship crossing the horizon will be burned to ashes by a ``firewall": a strong-energy momentum density in correspondence of the horizon, generated by quantum field theoretical effects.   

This conviction is supported by the firewall theorem \cite{Almheiri2013a}, which proofs the existence of such a strong energy-momentum tensor, on the basis of a few seemingly reasonable hypotheses.    In popular accounts (for a recent one, see for instance \cite{Wuthrich}), three assumptions are (erroneously, as we shall see) said to be proven incompatible: (i) unitarity of the quantum evolution, (ii) equivalence principle (absence of firewalls), and (iii) quantum field theory on curved spacetimes. The physical question, of course, is whether the (actual) hypotheses on which the theorem relies are realistic: whether they are reasonably implied by the physics we know.

Much of the discussion on the hypotheses of the theorem has focused on the unitarity of the asymptotic-time evolution.  Here I show that it is not unitarity the crucial suspect of the firewall theorem. There is a more subtle assumption which is the one that is likely to be unphysical.  This assumption is based on subtly mixing up of two different notions of entropy.  An analogous observation has been recently made in \cite{hotta}, where a model making the distinction concrete is presented\footnote{I was not aware of this reference when I posted the first version of this paper.}. Here I show that this distinction is actually forced by general relativity and quantum field theory.  \\

To start with, let us clear the discussion from a recurring confusion. The firewall theorem is based on considerations regarding conservation of information.  As repeatedly pointed out (see for instance \cite{Wallace2017}), there are two distinct issues regarding conservation of information: let me call them the ``long term" issue and the ``Page" issue:

\begin{figure}[b]
\includegraphics[width=6cm]{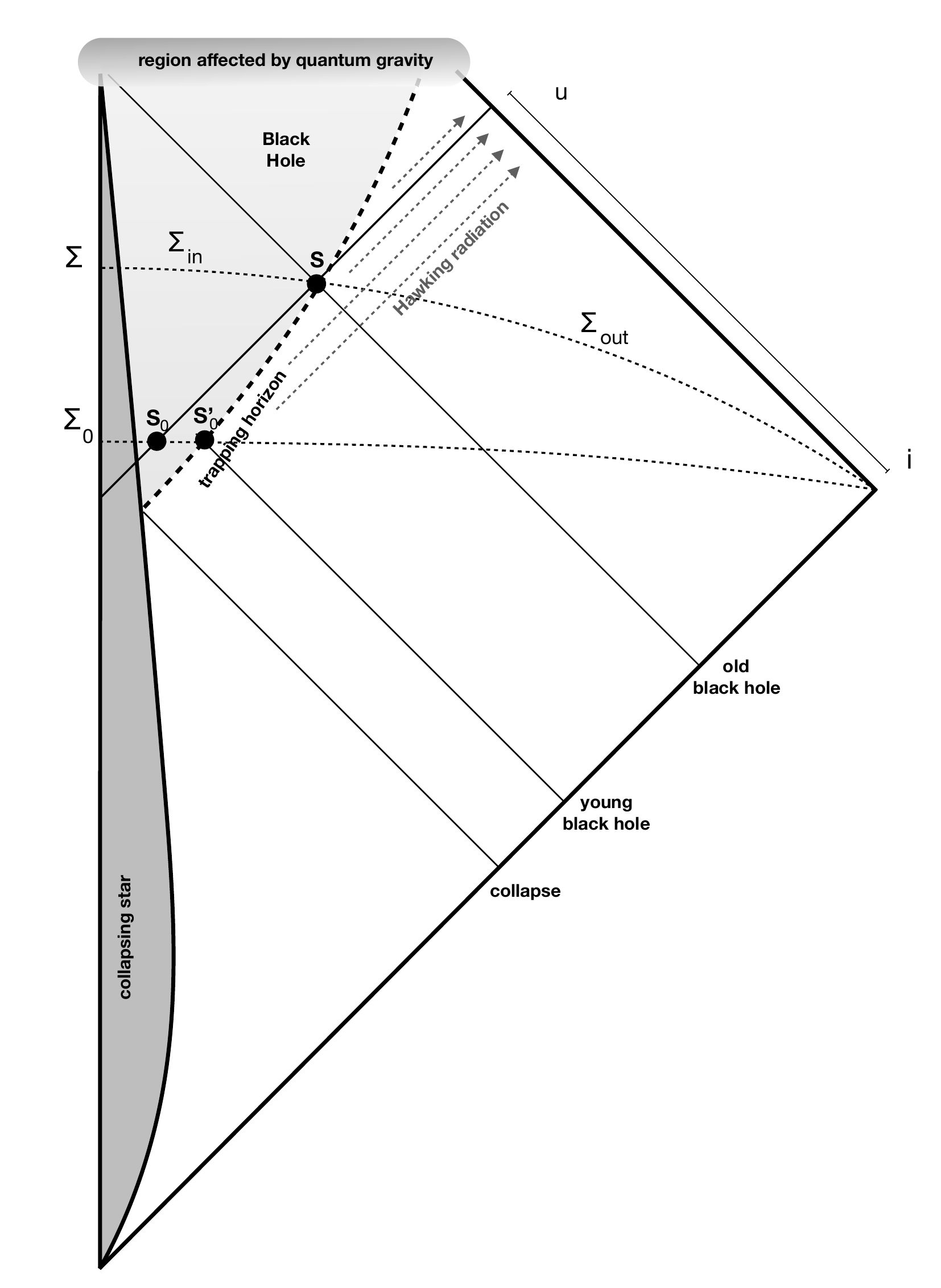}
\caption{\em The portion of spacetime relevant for the firewall issue. }
\end{figure}

(i)  The ``long term" issue is the observation that if there is a singularity in spacetime, as predicted by \emph{classical} general relativity (and perhaps corrected by quantum gravity), then part of the information on past null infinity falls into the singularity and does not reach future null infinity.  In this situation one may simply argue that there is no reason to expect unitary evolution between past null infinity and future null infinity, hence no reason to assume unitarity, which is one of the hypothesis of the firewall theorem.  Bob Wald and Bill Unruh, for instance, have eloquently made this point in \cite{Unruh2017b}.  This consideration is of course correct, but does not address the second version of the information issue, which is the one that has convinced many physicists that the unitarity hypothesis of the theorem is nevertheless reasonable.

(ii) The ``Page" version of the information issue \cite{Page1993c} regards \emph{only} the physics of a spacetime region \emph{before} the development of the singularity.  Also, it regards  \emph{only} the physics of a spacetime region \emph{before} the end of the black hole evaporation. That is, the ``Page" version of the information issue does not need to consider the regions (around the singularity and past the end of the evaporation) which are likely affected by quantum gravity. All considerations are restricted to a region where established theories are expected to be reliable.  Figure 1 is the Carter-Penrose diagram of a  spherically symmetric spacetime geometry around a collapsing star, on which there is an evolving quantum field $\phi$. The geometry takes into account the back-reaction of the Hawking radiation of the field.   The collapsed star generates a trapped region: the black hole. The boundary of this region is the (trapping) horizon.  In the limit in which we disregard the back-reaction of the Hawking radiation, the horizon is null, but taking  back-reaction into account, it is time-like.  The notion of \emph{event} horizon is not defined, because no assumption is made about the distant future, which depends on quantum gravity and is not relevant for the firewall theorem.  Considerations on this region are sufficient to derive the firewall theorem using Page's argument, hence the existence of a singularity is not relevant for the theorem.\\  

Let me review Page's key observation. Consider a sphere $S$ on the horizon at retarded time $u$ (see Figure 1).  The region of future null infinity preceding the time $u$ receives the Hawking radiation emitted until the horizon has reached $S$.   Consider the case where the black hole is ``old" at $S$, namely the area $A$ of $S$ is much smaller than the the initial horizon area $A_0$ at $S_0$.  The Hawking radiation arriving at future infinity around $u$ is in a mixed state.  If the initial state of the field was pure and evolution is unitary, this radiation must be correlated with something else: with what?   

There are two reasonable possibilities:  

(a) it is correlated with degrees of freedom inside the horizon; 

(b) it is correlated with degrees of freedom outside the horizon. In particular, late Hawking quanta may be correlated with early Hawking quanta. 

A part of the theoretical community has got convinced that the correct answer must be (b) because (a) is ruled out by the following argument by Don Page. Assume that:

{\em Assumption A:  The number $N$ of states of a black hole with which external degrees of freedom can be entangled at some given time is bounded by $N\sim e^{A/4}$ in Planck units ($\hbar=G=c=k=1$), where $A$ is the area of the horizon at that time.}

Then, as $A$ shrinks towards zero with time, so does $N$.  At some point necessarily $N$ is too small for the black hole to have enough states for the total entropy of the Hawking radiation to be accounted for by correlations with black hole states.    Hence (a) is ruled out. Hence (b) is the correct story: the late Hawking quanta must be correlated with the early Hawking quanta.

This argument is a key ingredient of the firewall theorem.  By an elegant-information theoretical argument, the theorem shows that a consequence of this is that the correlation across $S$ must be weak.  But quantum field theory tells us that all states with finite energy necessarily have strong short-scale correlations. The only possibility for having weak correlations is to go very far from a local vacuum, namely to have a strong local energy density. We refer the reader to the quantitative analysis of the firewall paper for the details.  The relevant point for us here is that the firewall theorem assumes crucially the Page argument above and, in turn, {\em Assumption A}. 

What are the arguments supporting {\em Assumption A}?   The simplest is the following: there is large evidence that a black hole of area $A$ has a (Bekenstein-Hawking) entropy $S=A/4$. Entropy counts the number of possible (orthogonal) quantum states: $S=\log N$ from which the assumption follows immediately.  

A more indirect argument, which relies upon confidence on tentative quantum gravity theories, is the actual black hole state counting in string theory \cite{Strominger1996} and in loop quantum gravity \cite{Rovelli:1996dv,Ashtekar:1997yu,Perez2017bbb}: both give a number of states which is finite and proportional to $e^{A/4}$. The convergence between the two rival current tentative theories of quantum gravity is remarkable, and provides credibility to the result.  Additional indirect evidence comes assuming the AdS/CFT duality \cite{Hanada2014}.
 \\

However,  {\em Assumption A} is the unphysical assumption of the theorem.  Why?  The arguments supporting it are well known and commonly accepted, they seem solid.  The subtle point is that there is a crucial illegitimate  step that must be taken to go from the Bekenstein-Hawking entropy and quantum-gravity state-counting (both of which I take here to be physically correct), to {\em Assumption A}.  The step is illegitimate   because it subtly confuses two distinct notions of entropy and two distinct notions of state.  \\ 

A first preliminary point is the observation that the use of entropy in Page's  Argument is not supported by the standard Bekenstein's entropy logic. The reason is the following. A  celebrated observation by Raphael Bousso is that the entropy bounds cannot be applied naively to volumes (here the hypersurface $\Sigma_{in}$ in Figure 1);  it can be shown to fail in this form. Rather, it must be applied covariantly to null surfaces \cite{Bousso2002a}. To apply Busso's covariant version of the entropy bound to a null surface, this must be a Light-Sheet, in the terminology of \cite{Bousso2002a}, namely its expansion must be negative moving away from the surfaces and the null surface must close.  The problem is that \emph{there is no Light-Sheet for surface $S$}.  The four null surfaces emanating from a sphere $S$ on horizon of an old black hole are depicted in Figure 1.  Three of these are expanding and the fourth does not close in the region where quantum gravity can be  neglected.  The crucial one is the internal past light surface, namely the lower left one in the Figure:  naively, one may think that this has negative expansion when moving away from $S$, but that's not true. This is because the area of the sphere $S_0$, situated in the internal past light cone of $S$ is exponentially close to the area of the sphere $S_0'$, situated on the trapping horizon, which is much bigger that the area of $S$, precisely because the black hole is evaporating.  Therefore: \emph{the entropy bound does not apply after the black hole has shrunk}. (Nor is the quantum version of the entropy bound, as  mentioned in \cite{Bousso2016}.) This implies that all the standard arguments showing the maximal entropy falling into a region is bound by the area {start failing as soon as the horizon surface shrinks}. \\

Let me now come to the main point: {\em Assumption A} follows from a subtle confusion between two distinct notions of state and two distinct notion of entropy. The problem is the ambiguity in the notion ``number of states of a black hole".  This phrase can have two distinct meanings, and the confusion between them is the origin of the troubles. To clarify this point, consider the following. 

The ``states" of a classical (or quantum) physical systems are the sets of values (or  probability distributions of the values) that the observables of a system can take.\footnote{Formally, a state is a function on the space of the observables.  If $\Gamma$ is the classical phase space and $A:\Gamma\to R$ an observable, then a pure state $s\in\Gamma$ defines the (Gelfand) function $s(f)=f(s)$ on the space of the observables, and so does a mixed state $\rho:{\Gamma}\to R^+$ via $\rho(A)=\int_\Gamma\rho A$.  If $\cal H$ is the Hilbert space of a quantum system, and the self adjoint operator $A:{\cal H}\to {\cal H}$ is an observable, then a pure state $\psi\in\cal H$ defines a function $\psi(A)=\langle \psi |A| \psi \rangle$ on the algebra of the observables and so does a mixed state $\rho:{\cal H}\to\cal H$ via $\rho(A)=Tr[\rho A]$.} The point is that the notion of ``state" \emph{depends} on the algebra of observables for which it is defined. 

Consider a physical system confined inside a finite box. Suppose this system has two kinds of degrees of freedom. A set of degrees of freedom $q$ that can interact with the exterior of the box and a set of degrees of freedom $x$ that for a finite but long time interval $T$ are isolated for all practical purposes.  Let ${\cH}^{(q)}$ be the Hilbert space of the first, ${\cH}^{(x)}$  the Hilbert space of the second, and assume that both are finite dimensional.  Suppose we are outside the box and interact with the box during the time interval $T$. What is the relevant maximal thermodynamical entropy $S^{\rm therm}$ describing the thermal behaviour of the box? 

The answer is $S^{\rm therm}=\dim {\cH}^{(q)}$, because entropy governs the heat and energy exchanges of a system with the exterior; if the  $x$ degrees of freedom are decoupled, their value cannot change in the interaction, therefore they do not partecipate in these exchanges. Energy never thermalises to them. Therefore they are irrelevant for the thermal behaviour of the box. Therefore they do not contribute to $S^{\rm therm}$

Now let us ask a different question.  If the entire box and its exterior are in a pure state, what is the maximum entanglement entropy $S^{\rm ent}$ between the exterior of the box and the interior? Now there is no reason not to include the $x$ degrees of freedom in the counting, because the full state may be in a quantum superposition including different values of these variables, for instance established before the interval $T$ during which these degrees of freedom are effectively decoupled.  Hence 
\be
S^{\rm ent}=\dim (\cH_q\otimes \cH_x)>\dim (\cH_q)=S^{\rm therm}.
\ee
Therefore: \emph{entanglement entropy can be larger than thermodynamical entropy}.  \\ 

Let us see how these considerations play out in the black hole case. There is evidence that the thermal interactions of a black hole with horizon area $A$ are governed by a thermodynamical entropy $S^{\rm therm}=A/4$. I assume here, as commonly done, that this is the case.  Is the entanglement entropy of a black hole bounded by its thermodynamical entropy? 
 
For a young black hole, the Bousso bound gives us a positive answer, because before any evaporation the interior past light cone of a sphere on the horizon has a past Light-sheet and the bound applies. Since the entropy that may have crossed the Light-Sheet is bounded, so is the maximum number of possible interior states that may have had the chance of getting entangled with the exterior.  

But after the back-reaction leading to the evaporation starts, the Bousso entropy bound cannot be invoked anymore.  The Area decreases: does the maximal entanglement decrease as well? 

It is shown in \cite{Rovelli2017a} that this is impossible. I review the argument here.  Consider two Cauchy surfaces $\Sigma_0$ and $\Sigma$ that are part of a foliation of spacetime as in Figure 1. Let $S$ and $S_0'$ be the intersections of these surfaces with the horizon.  Let $S_0$ be the intersection of $\Sigma_0$ with the interior past light cone of $S$.  Let  $\Sigma^\pm_0$ and $\Sigma^\pm$ be the portions of these surfaces inside and outside $S_0$ and $S$ respectively.  Call  $\cA_0$ and, respectively, $\cA$, the algebra of the local quantum field theoretical observables on $\Sigma^-_0$ and $\Sigma^-$   and the corresponding Hilbert spaces  $\cH$ and $\cH_0$. Since $\Sigma_0^-$ is in the past causal domain of $\Sigma^-$, relativistic dynamics demands that $\cA_0^-$ is a subalgebra of $\cA^-$. Therefore any state on the second is also a state on the first, and therefore $\cH_0$ is a subspace of $\cH$. Either these state spaces are infinite dimensional, or they are finite dimensional. If they are infinite dimensional, the entanglement entropy can be as large as one wishes, and {\em Assumption A} is false. If they are finite dimensional, the fact that one is a subspace of the other implies that $\dim \cH_0 \le\dim \cH$. Therefore the entaglment entropies on the two surfaces are related by
\be
S_0^{\rm ent}=\dim \cH_0\le\dim \cH=S^{\rm ent}
\ee
But $A_0>A$ because of the backreaction of the Hawking radiation and the fact that the area of $S_0$ is exponentially close to the area of $S_0'$ implies that  $A_0 > A$. Assuming that the Bekenstein-Hawking entropy is the correct thermodynamical entropy we have
\be
S_0^{\rm therm}=A_0/4>A/4=S^{\rm therm}
\ee
The last two equations imply that for a black hole necessarily 
\be
S^{\rm ent}\ne S^{\rm therm}. 
\ee
PageÕs argument relies on interpreting $S^{\rm therm}=A/4$ as a standard density of states for the black hole. This is wrong, as shown above.  

This, by the way, is precisely what naive intuition suggests: the thermodynamical entropy $S^{\rm therm}$ counts the quantum degrees of freedom \emph{on the horizon}, which are responsible for the thermodynamical behaviour of the black hole. While the entanglement entropy $S^{\rm ent}$ counts the quantum degrees of freedom \emph{inside the black hole}.  The first decreases as the horizon shrinks.  The second increases because as more information falls in to the black hole, more entanglement is possible.  

Importantly, recall that according to classical general relativity, which is assumed valid in this regime, the volume inside the black hole ---defined as the surface of maximal area bound by a sphere on the horizon--- \emph{increases} even as the horizon area shrinks \cite{Christodoulou2015}. On this foliation the geometry of an evaporating black hole is that of an increasing volume closed by decreasing throat. \\ 

At first sight this conclusion may seem to be in conflict with results in string theory, in the context of gauge/gravity duality models and in loop quantum gravity. But at a closer inspection, this is not the case.  The Strominger-Vafa-style counting of D-brane bound states \cite{Strominger1996} identifies the $S^{\rm therm}$ with the log of the number of string states  and in the gauge/gravity duality the logarithm of the boundary gauge-theory partition function matches the  free energy $F = E - T S^{\rm therm}$ of bulk black holes \cite{Hanada2014}.  However, these calculations are in the limit when there is an event horizon: both string theory and gravity/gauge duality describe the bulk physics in terms of asymptotic observables and the calculations are made under the hypothesis of the existence of an event horizon.  But this is not relevant to describe the physical situation in which quantum gravity may correct the dynamics in the long term future, preventing the formation of the internal singularity and the event horizon. Results for this case, which is the physically relevant one, are not available.   Quite to the contrary,  results in lower dimension \cite{Fitzpatrick2016} show that the AdS/CFT correspondence gives the internal classical geometry with singularity and event horizon only in the (unphysical) ``classical" $N\to\infty$ limit; but in this limit, which is singular, unitarity is lost, while unitarity holds \emph{before} this unphysical limit.   String and AdS/CFT state counting, therefore, refer to a notion of state that is well defined, but is not the one relevant for the entanglement entropy. 

Loop quantum gravity state counting makes the point even more transparent: the result agrees with string theory but the states counted (in the various version of the loop quantum gravity derivation of the Bekenstein-Hawking formula  \cite{Rovelli:1996dv,Ashtekar:1997yu,Perez2017bbb})  are always explicitly \emph{horizon} states. These are the ones that govern the thermodynamical behavious of the hole and they are \emph{less} than the total number of internal states where information may have fallen. The information fallen inside is lost to the outside observers only in the approximation in which the trapping horizon is treated as an event horizon. 

All this shows clearly that there is no compelling argument for believing that the late Hawking quanta are entangled with the early ones. The interior degrees of freedom of the black do not contribute to its Bekenstein-Hawking entropy, do not contribute to Strominger-Vafa style counting (which is not relevant for trapping horizons)  but they can nevertheless purify the Hawking radiation also when the black hole is old. Therefore a key hypothesis of the firewall theorem falls. 

\begin{figure}[t]
\includegraphics[width=8cm]{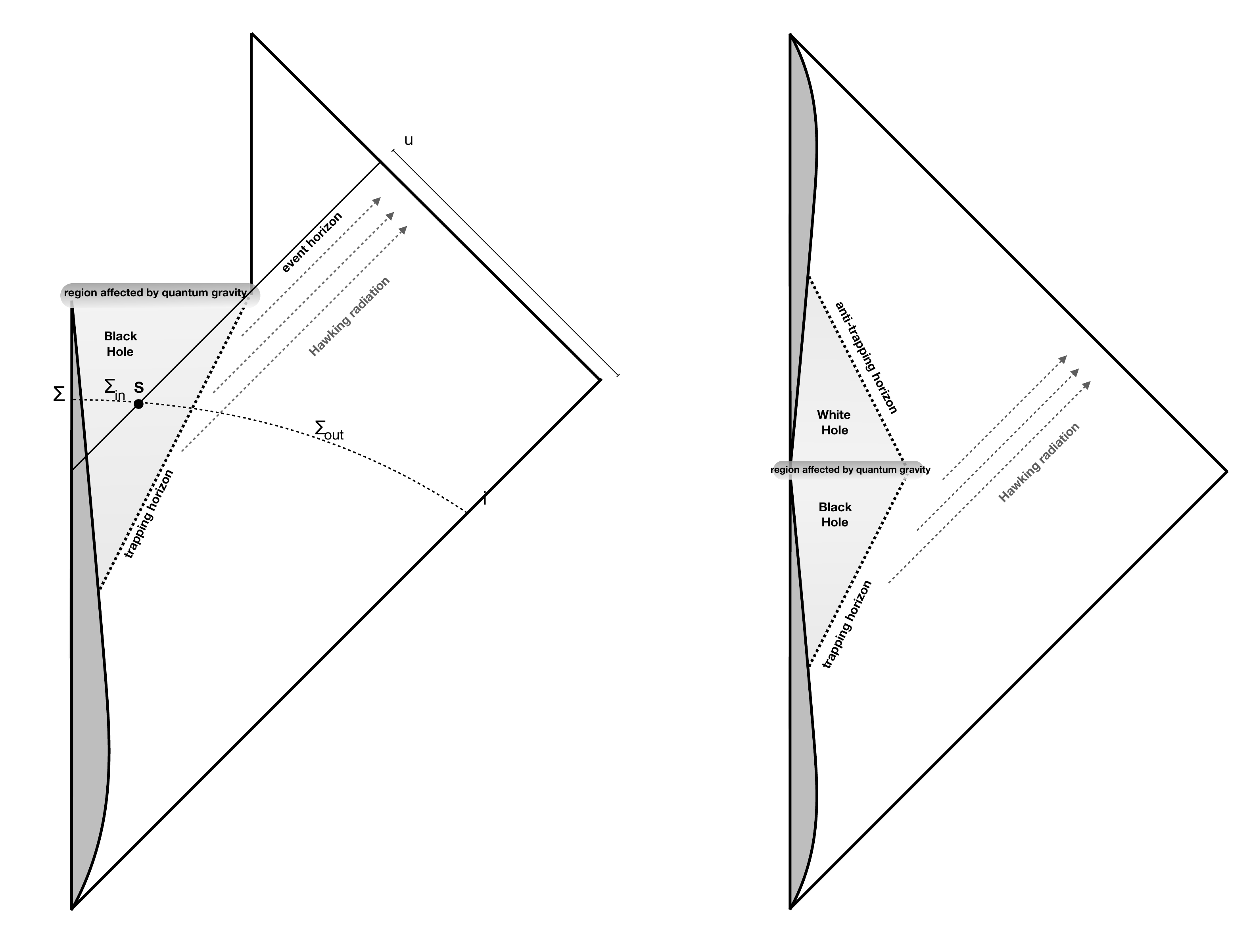}
\caption{\em Left: Popular but unlikely scenario for the end of black hole evaporation. Right: A more reasonable hypothesis about the (average) spacetime geometry.}
\end{figure}

What remains alive, is the naive version (i) of the information issue. This, however, involves the distant future of the hole, which is affected by quantum gravitational phenomena.   If the firewall theorem depends on credences about what quantum gravity implies, it looses its teeth. \\ 

Let me also briefly discuss for completness some possibilities about the distant future, although this is not necessary for the main point of this paper.  A common popular hypothesis in the literature is that at the end of the Hawking evaporation a black hole simply disappears from the universe, popping into nothing.  This scenario is represented by the popular left spacetime diagram in Figure 2. 

It is important to emphasise that nothing in current physics really implies that this is the geometry of a fully evaporating black hole: when the area of the black hole becomes very small, we are deeply into the quantum gravitational regime.  The idea that the black hole may disappears ``just because it is small" is ungrounded and superficial, especially because its internal volume (defined above) remains big \cite{Christodoulou2016a}. 

Still: what would be the behavior of the quantum field if this was the relevant effective geometry for the quantum field theory? The answer is obviously the one by Unruh and Wald in \cite{Unruh2017b}. There is no reason for unitarity in the context of quantum field theory on a background, \emph{if} this is the spacetime geometry.  A quantum theory of gravity where there is a unitary map from past to future null infinity is a theory of quantum gravity where the classical geometry of the left panel of Figure 2 is corrected by quantum gravitational phenomena. Those that expect that there is unitarity at infinity \emph{and also} that the left diagram of Figure 2 represents the exact causal structure of spacetime are simply holding contradictory assumptions. The AdS/CFT results mentioned above  \cite{Fitzpatrick2016} offer evidence in favour of this point.  

A more likely (effective) geometry that quantum gravity may imply is depicted on the right side of Figure 2.  This has been recently  extensively explored \cite{Bianchi2018b}.  The singularity at zero Schwarzschild radius is avoided by quantum gravity: spacetime continues into a white hole geometry. This is predicted explicitly by loop quantum gravity \cite{Ashtekar2018c}. After the Hawking radiation has shrunk the horizon, quantum gravity allows the trapping horizon to tunnel into an anti-trapping horizon, which bounds the white hole produced in the interior \cite{Christodoulou2016}.  Resulting Planck size white holes are long-living remnants that are protected by causality from being spontaneously produced in low-energy physics and are stabilised by their small throat and large interior volume \cite{Rovelli2018f}. The information that falls into the black hole slowly leaks out of the long-living small white hole, constraining its life to be long  \cite{Bianchi2018b}.   Phenomenological consequences of the existence of similar \cite{Barrau2016c} or related \cite{Barrau2014b} remnants of primordial black holes, including the possibility that they form a component of dark matter, are being explored.  

Remnants of this kind are sometimes said to be incompatible with AdS/CFT (see for instance \cite{Almheiri2013b}), because they demand too many states at too little energy. This is indeed the way the firewall theorem is presented for instance in \cite{Marolf2013,Almheiri2013b}.  But this is a far cry from the derivation of a theorem from simple seemly reasonable hypotheses.  If one includes the precise physical validity of some AdS/CFT model among its hypotheses, the theorem  looses its strength as a convincing argument for the real existence of firewalls.  In the eyes of most physicists, it rather becomes a reduction of this version of AdS/CFT {\em ad absurdum}.   As mentioned, alternative are possible: the AdS/CFT correspondence itself might give the standard black hole counting in the unphysical ``classical" $N\to\infty$ limit where an event horizon exist but  unitarity is lost; while unitarity holds before the unphsyical limit is taken \cite{Fitzpatrick2016}: an interesting possibility which again takes us outside the hypotheses of the theorem.    Still another possibility is that the information is taken by  correlations with fundamental pre-geometric structures \cite{Perez2015b}.

The physics that we consider more reliable, which include quantum field theory on curved spacetime in the region where quantum gravity is expected to be irrelevant, predicts that there are no firewalls: the evolution from a vacuum state on past null infinity can be computed using standard quantum field theory  methods and is regular on the horizon.  If a speculative hypothesis contradicts this reliable prediction, the likely solution is that the speculative hypothesis is wrong.  

These final considerations are not relevant for the main conclusion about the firewall theorem presented in this paper: the popular idea that unitarity and the validity of quantum field theory in low-curvature region are sufficient to prove firewalls (``breaking the equivalence principle") is, simply, wrong.  {\em Assumption A} is not supported by these hypotheses and is grounded on a subtle confusion between thermodynamical entropy and entamglemengt entropy; as soon as this assumption falls, the theorem falls.  The theorem does not rely on a set of hypothesis simple enough to make its conclusion plausible. 

If you consider crossing the horizon of a black hole with a starship, you should have other concerns than been burned to ashes at the horizon. \\

\centerline{---}
\vspace{.3cm}

I thank Don Marlof, Raphael Bousso, Eugenio Bianchi and Simone Speziale for reading and commenting preliminary version of this note and for very useful inputs. 

\bibliographystyle{utcaps}
\bibliography{/Users/carlorovelli/Documents/library}
\end{document}